# Active Rosensweig Patterns

Carlo Rigoni[1,2,3,*], Max Philipp Holl[2,4], Alberto Scacchi[5,6], Emil Stråka[2,4], Fereshteh Sohrabi[1,2], Mikko P. Haataja[7], Maria Sammalkorpi[2,4], Jaakko V.I. Timonen[1,2,*]

[1] Department of Applied Physics, Aalto University, Konemiehentie 1, 02150 Espoo, Finland

[2] Academy of Finland Center of Excellence in Life-Inspired Hybrid Materials (LIBER), Aalto University, Finland.

[3] Institute of Science and Technology Austria, Am Campus 1, 3400 Klosterneuburg, Austria

[4] Department of Chemistry and Materials Science, Aalto University, Kemistintie 1, 02150 Espoo, Finland

[5] Department of Bioproducts and Biosystems, Aalto University, Kemistintie 1, 02150 Espoo, Finland

[6] Department of Mechanical and Materials Engineering, University of Turku, Vesilinnantie 5, 20014 Turku, Finland

[7] Department of Mechanical and Aerospace Engineering, and Princeton Materials Institute, Princeton University, Princeton, New Jersey 08544, United States

\* Corresponding author; E-mail: carlo.rigoni@ist.ac.at, jaakko.timonen@aalto.fi

**Abstract**

**Ferrofluids, colloidal dispersions of magnetic nanoparticles, are renowned for pattern formation like few other materials. The Rosensweig instability of a horizontal ferrofluid-air interface in perpendicular magnetic field is especially well known classically, this instability sets the air-ferrofluid interface into an array of spikes that correspond to a new free energy minimum of the system. However, once the pattern is formed, it does not exhibit any notable thermal or non-equilibrium fluctuations, i.e., it is passive. In this work, we present an active version of the Rosensweig patterns. We realize them experimentally by driving a dispersion of magnetic nanoparticles with an electric field into a non-equilibrium gradient state and by inducing the instability using a magnetic field. The coupling of electric and magnetic forcing leads to patterns that can be adjusted from quiescent classic Rosensweig-like behavior (at low activity) to highly dynamic ones displaying peak and defect dynamics, as well as tunability of structure periodicities beyond what is possible in the classic systems (at high activity). We analyze the results using an active agent-based approach as well as a continuum perspective. We construct a simple equilibrium-like effective Rosensweig model to describe the onset of the patterns and propose a minimal Swift-Hohenberg type model capturing the essential active pattern dynamics. Our results suggest that classic continuum systems exhibiting pattern formation can be activated to display life-inspired non-equilibrium phenomena.**

# Introduction

The Rosensweig instability is one of the cornerstones of pattern formation in both classic[1-3] and even quantum fluids[4]. In its classic manifestation occurring at a horizontal interface between a magnetic and a non-magnetic fluid in a perpendicular magnetic field, this instability leads to formation of a hexagonal lattice of spikes (Fig. 1a). The pattern formation is driven by the reduction of magnetostatic energy at the expense of increasing gravitational and interfacial energies[2]. Square lattices are also possible, as demonstrated both experimentally[5,6] and theoretically[7,8]. The evolution of the onset of the Rosensweig instability has been quantitatively explored both with time resolved experiments[9,10] and through a careful theoretical analysis, both in the linear and non-linear regimes[3]. Pattern geometry also depends on the vertical and horizontal confinement[11,12]. The Rosensweig instability has been also studied in droplets[13] and soliton spikes have been discovered[14]. In addition, there have been studies coupling the Rosensweig instability to other phenomena and dynamic fields, such as AC fields[13] and standing waves[15,16]. Importantly, because the core pattern formation is driven by energy minimization, and because most experimental systems are essentially athermal, the resulting patterns remained stationary once formed.

In this work, we demonstrate active Rosensweig patterns. Our platform is based on electrically driven magnetization gradients in a magnetic fluid subjected to an external magnetic field[17,18]. We show that the applied voltage and current create a nanoparticle gradient with fluctuations that affect the pattern formation dynamics, leading to rich active dynamics including continuous translation, merging and splitting of peaks, and collective motion. A simple theoretical model is developed to rationalize the observed pattern wavelength and time scales near the onset of the instability. A minimal model based on two non-variationally coupled Swift–Hohenberg equations is put forward to capture the active dynamics of the system in the nonlinear regime. However, we find, by extensive scanning of the parameter space, that this model fails to fully capture the dynamic features and that a more phenomenological random advection Swift–Hohenberg model is in better quantitative agreement with experimental observations.

# Experimental system for active Rosensweig patterns

The electroferrofluid consists of a mixture of n-dodecane with 300 mM of sodium bis (2-ethylhexyl) sulfosuccinate (AOT) and oleic acid coated iron oxide nanoparticles (NPs) synthesized in house (see Methods and Fig. S1 of the supplementary information). A fraction of the ferromagnetic NPs in the electroferrofluid are charged negatively through the interaction with the AOT inverse micelles present in the mixture[17], enabling control of the local NP concentration $c(x, y, z)$ using both electric and magnetic fields (Fig. 1b). In a Hele-Shaw microelectrode cell (Fig. S2), applying a voltage allows transition from the equilibrium NP concentration $c_0$ (Fig. 1c, left) to a steady state dissipative gradient of NPs (Fig. 1c, center). Applying a magnetic field perpendicular to the electrodes induces the formation of a pattern of dark and light spots in the microscopy images corresponding to regions of higher and lower concentration of NPs (Fig. 1c, right) resembling the Rosensweig pattern in ferrofluids at ultralow interfacial tension[6].

For a thorough description of the behavior of the electroferrofluid in this specific setting, we first characterize the phase behavior as a function of its two main external "knobs": the voltage $U$ and the external magnetic field $B = \mu_0 H_0$ measured in units of T with $\mu_0$ being the vacuum magnetic permeability and $H_0$ the external applied magnetic field measured in units of A/m. The phase diagram is characterized by three states: the presence of steady-state gradient, magnetic pattern, and magnetic NP aggregates (Fig. 1d). At low values of $U$ and $B$, only the NP gradient state is observed, and therefore the top view microscopy images show no changes. The boundary between the gradient-only region and the pattern formation region suggests that the critical value of the magnetic field to observe the instability, $B_c$, is voltage dependent. At high $B > 10$ mT, the dipole-dipole magnetic attraction overcomes the electrostatic repulsion between the negatively charged NPs, and reversible aggregates of average size around 1 μm form in the sample and vanish in short time after the $B$ field is switched off. The steady passive limit structure of the pattern is a square lattice, and, in addition, the pattern fluctuates over time exhibiting active behavior with translation (Fig. 1e), and merging and splitting of peaks (Fig. 1g). The intensity of the pattern activity grows as a function of the voltage.

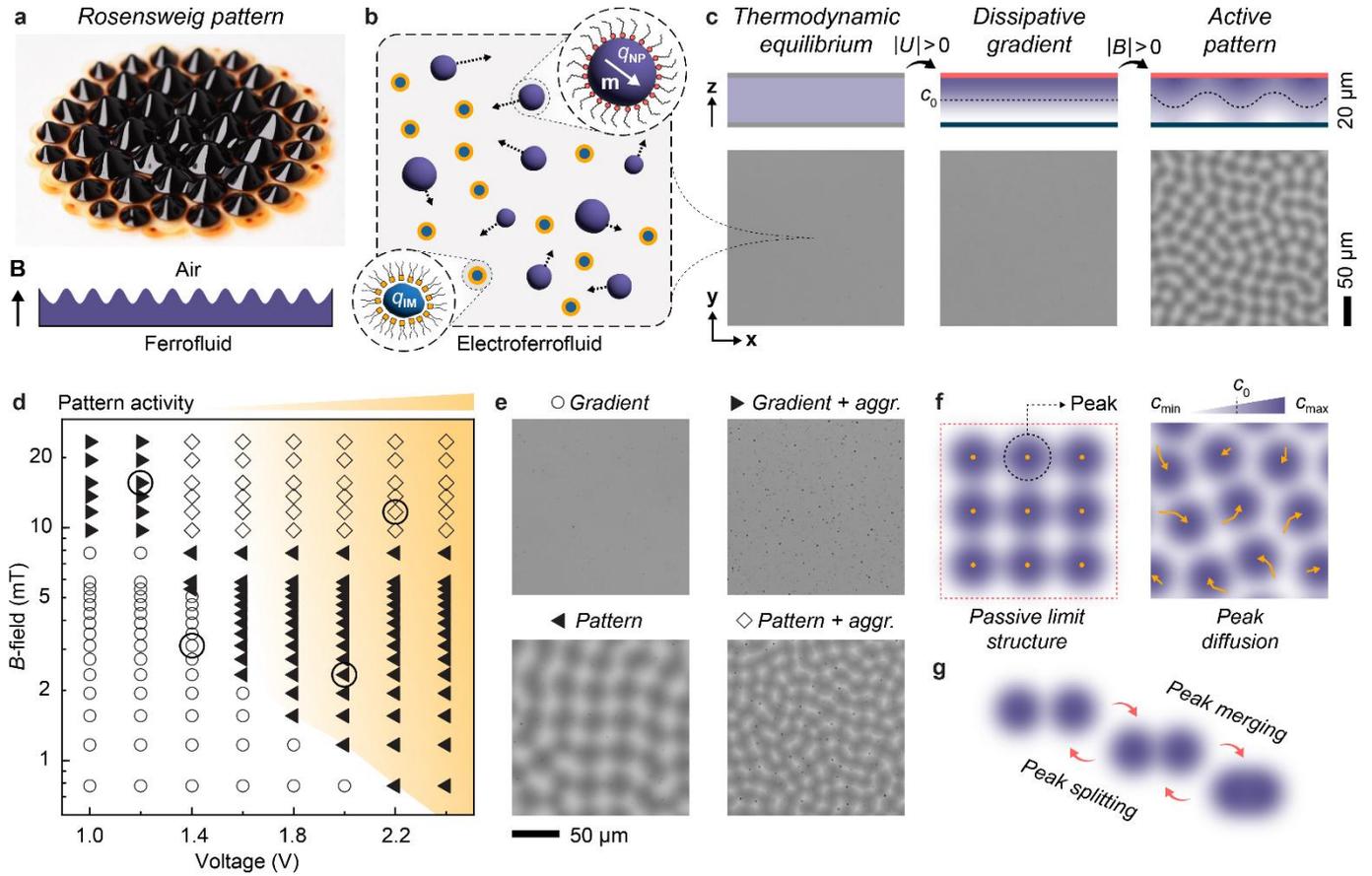

**Fig. 1 | From classic to active Rosensweig patterns. a.** Photo and concept scheme of the classical Rosensweig instability. Figure adapted and reprinted from ref.[17] **b.** Schematic representation of the composition of the electroferrofluid at the nanoscopic scale. **c.** Scheme of the distribution of the nanoparticles in the cell (top row) and corresponding microscopy image taken (bottom row) of the different states of the system leading to the active pattern. **d.** Phase diagram of the different states achievable in function of the voltage and the magnetic field. **e.** representative microscopy images for each state. **f.** Schematic representation of the difference between the passive limit square geometry pattern and the active pattern where continuous peak diffusion takes place. **g.** Schematic representation of the peak merging and splitting happening in the active pattern.

## Formation dynamics of the active Rosensweig patterns

The patterns exhibit a complex time evolution behavior (Movies 1, 2). The time evolution has two different time scales (Fig. 2a): fast onset dynamics (Movie 1) and slow annealing dynamics (Movie 2). The onset of the pattern formation generally occurs within a few minutes starting with the initial displacement of the uniform distribution of NPs and reaching an apparent steady state for the pattern periodicity and amplitude (Fig. 2b). The voltage clearly influences the dynamics and therefore can be used to tune the speed of the onset process. More specifically, the onset dynamics are characterized by exponential laws and both the timescale for the setting of the periodicity $\tau_p$ and the timescale for the growth of the amplitude $\tau_a$ decrease with increasing voltage (Fig. 2c). Within a simple model for the charge transport kinetics through the cell (Supplementary Note 1) that considers the interaction between the charged AOT inverse micelles and the NPs, one arrives at a linear steady-state NP concentration profile given by $c(z) = c_0 + zJ/D_{NP}$ where $J$ is the current density, $D_{NP}$ is the diffusion coefficient of the NPs and $z$ the distance measured from the mid-plane of the cell. The model also predicts the formation timescale to depend on the current density as $\tau_{p,a} \sim J^{-3}$. Current density in our system scales as[17] $J \sim (U - U_0)$, where $U_0$ is a threshold voltage above which the current increases linearly. Hence $\tau_{p,a} \sim (U - U_0)^{-3}$ is in good agreement with the experiments (Fig. 2c).

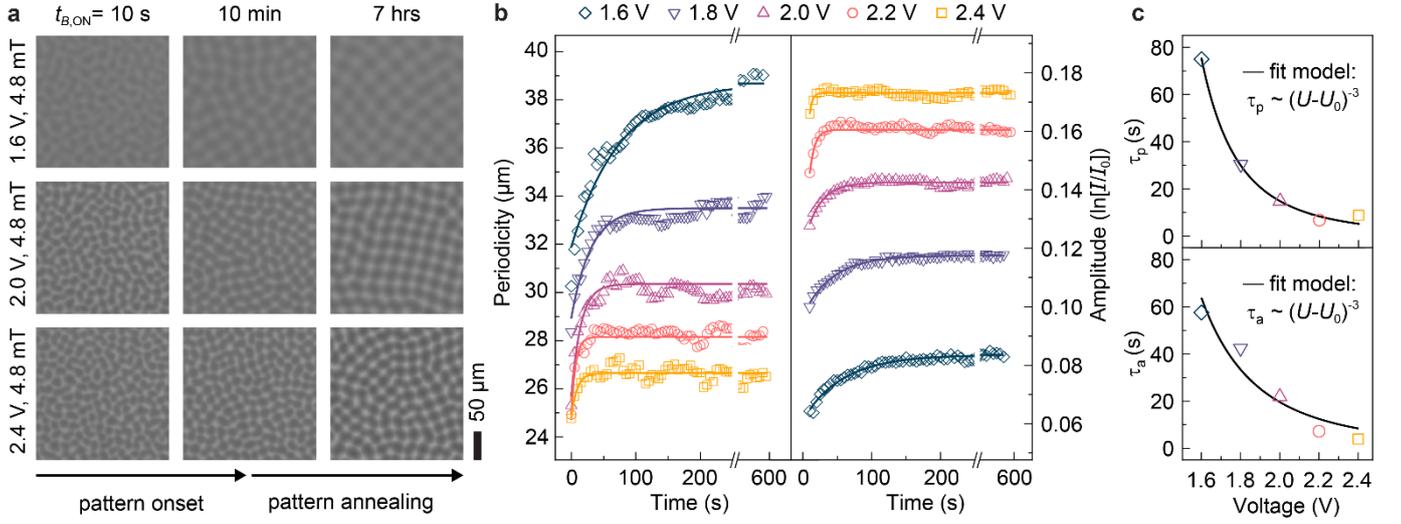

**Fig. 2 | Formation dynamics. a.** Snapshots of the different states observed in the system in function of time for three different voltages. **b.** Periodicity and amplitude of the pattern as a function of time and of the voltage applied. **c.** Timescale values $\tau_{p,a}$ as a function of the voltage.

## Average periodicity and amplitude of the active Rosensweig patterns

The study of the periodicity (Fig. 3a) and the amplitude (Fig. 3b) reveals that the voltage strongly influences the geometrical properties of the pattern. The pattern periodicity, obtained by measuring the peak-to-peak distance of the concentration pattern, confirmed by fast Fourier transform (FFT) analysis (Fig. S3), decreases as a function of both $U$ and $B$ with an apparent initial maximum value at low $U$ and $B$ close to 50 μm (Fig. 3a). This behavior is notably richer than that observed for the Rosensweig instability in regular ferrofluid systems, where the periodicity is essentially constant as a function of the applied magnetic field[3]. Combination of a simple analytical model and particle-based simulations (see Supplementary Notes 1 and 2, Fig. S4) suggests that the pattern periodicity scales with the voltage and magnetic field as $\lambda \sim \frac{1+B_0^2/B^{*2}}{JB_0^2}$, where $B_0$ is the external applied magnetic field and $B^*$ a characteristic magnetic field with magnitude on the order of $B^* \approx 30$ mT, where $B^* = \sqrt{\mu_0 k_B T/(c_0 \alpha^2)}$ with $k_B$ being the Boltzmann constant, $T$ the temperature, and $\alpha = \chi/c_{NP}$ with $\chi$ the magnetic susceptibility (see Supplementary Note 1 for additional details on the derivation of $B^*$). Even though the scaling approach is only valid close to the onset of instability, the model predictions remain surprisingly robust far from it. An exception is the pattern periodicity data at low $B$ that reaches a maximum value of 50 μm in the experiment, while the simple model fails to capture such a maximum. Nonetheless, the main prediction of the model is that the periodicity data should collapse on a single curve that is dependent only on the magnetic field magnitude, if multiplied by $(U - U_0)$. This is indeed the case for $U_0 = 0.2803$ V (Fig. 3a inset, Fig. S5). Furthermore, the model predicts that these collapsed data should reach an asymptotic value at high $B_0$, which is qualitatively visible from the plotted graphs (Fig. 3a inset). Finally, to follow the theory, an approximate model with a dependency of $\sim B_0^{-2}$ can be matched to the experimental data (Fig. 3a inset, theory match). The best fit of $B_0^a$ left free to change sets the exponent on a value of $a = -0.36$ and acceptable values of $a$ as judged by visual inspection span the range $-3 < a < -0.25$.

The pattern amplitude dependence on $U$ and $B$ is even more complicated than that observed for the periodicity. The values of relative concentration measured at the peak locations by the Beer-Lambert law ($\Delta c \sim -\ln[I/I_0]$) employed to quantify the amplitude (Fig. S6), are non-monotonic as a function of $B$, with a peak at low $B$ values while consistently growing as a function of $U$ (Fig. 3b). In this, $I$ is the light intensity at the selected location with $U \neq 0$ and $B \neq 0$, and $I_0$ is the light intensity at the selected location with $U = 0$ and $B = 0$. Below 1.4 V, it is not possible to observe pattern formation when the magnetic field is applied. The behavior at the lowest voltage at which pattern formation was observed (1.4 V) resembles the amplitude profiles for the classical Rosensweig instability: a fast initial rise followed by a slow increase as a function of the magnetic field[3]. A peak in the values emerges with increasing $U$ and is highly asymmetric with a sharp increase followed by a slow decay. This non-monotonic behavior introduced by the voltage is another sign of a nontrivial coupling between the magnetic and electric fields. Both the maximum value of the peak and the asymptotic value at large $B$ depend linearly on $U$ (Fig. 3b, inset). No sign of hysteresis was detected in the periodicity, while the amplitude data presented a hysteresis only between increasing and decreasing $B$, in accordance with previous literature[3] (Fig. S7).

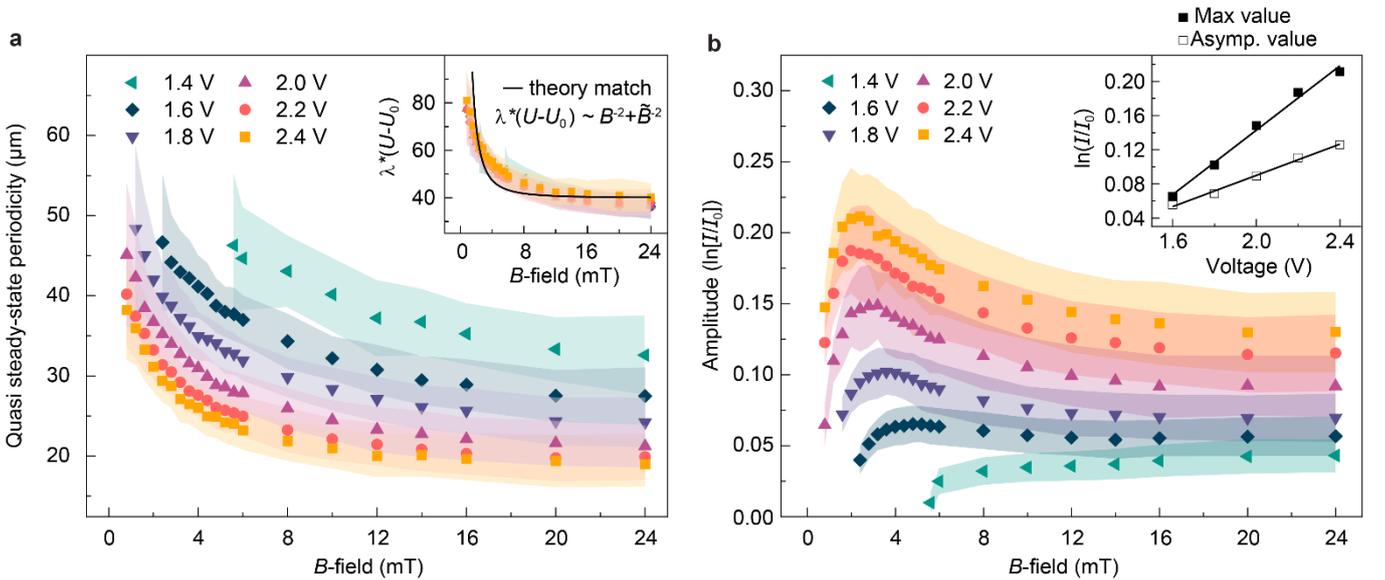

**Fig. 3 | Early-time pattern periodicity and amplitude. a.** Periodicity of the pattern as a function of $U$ and $B$ (10 min after the fields are changed). The value corresponds to the average distance between the peaks and the colored error band corresponds to the standard deviation. In the inset the same data are plotted rescaled with $(U - U_0)$ and $U_0 = 0.2803$ V. **b.** Equivalent amplitude of the pattern estimated by light absorption as a function $U$ and $B$ (10 min after the fields are changed). The value corresponds to the average of all values of the maxima in the images and the colored error band corresponds to the standard deviation. In the inset, the maximum value of the peak and the asymptotic value of the curve at high $B$ are plotted.

**Elementary dynamic processes of the active Rosensweig patterns**

Following the pattern onset behavior, a slow process of annealing begins, occurring over several hours. A relative neighborhood graph of the pattern "valley" positions is employed to monitor the annealing and the activity dynamics (Fig. 4a, Movie 3). The reconstruction of the geometry reveals that the preferred arrangement is a square lattice with a frequent presence of triangular defects. The square geometry arises naturally in the Rosensweig instability under extreme conditions, such as a very high magnetic field in combination with a high magnetic susceptibility[7,19], or an ultra-low interfacial tension in combination with a low magnetic susceptibility[6,7]. While the expected annealing process seems to occur for the lower voltages (take for example the case at 1.6 V, Movie 4), the system instead settles into an apparent steady state at 2.4 V, behaving as an active pattern (Movie 5) displaying time-invariant defect distribution. The signature phenomena of activity in the system over time are continuous translation of the peaks in space, merging of neighboring peaks and splitting of parent peaks into two child peaks (Fig. 4b).

The effects of the activity can be quantified via inspection of four different parameters. The patch size (correlation length) within which the pattern is consistent (quantified as the inverse of the full width at half maximum (FWHM) of the FFT peak over time) and normalized by the pattern periodicity (Fig. 4c) reveals again that while for 1.6 V and 2.0 V the pattern correlation length progressively increases to a larger number of period lengths, for 2.4 V the normalized patch size seems to be constant over time. Similarly, the standard deviation of the distribution of the angles in the pattern ($\sigma_\theta$) decreases over time for 1.6 V and 2.0 V but remains constant for 2.4 V (Fig. 4d). Additional evidence of this high voltage response can be seen in the decrease of the relative number of triangular defects over time for the different voltages: at the highest voltage more defects exist, and their total number remains constant (Fig. 4e). Finally, tracking the peaks in time (Fig. S8 and Movie 6) allows to measure their lifetime. In particular, analyzing the lifetime of the peaks present in the last 30 minutes of the experiments reveals that while many peaks at 1.6 V and 2.0 V surpass even 7 hours lifetime, at 2.4 V the vast majority of peaks last less than 1 hours. Furthermore, no peak survives for the entirety of the experiment (Fig. 4f).

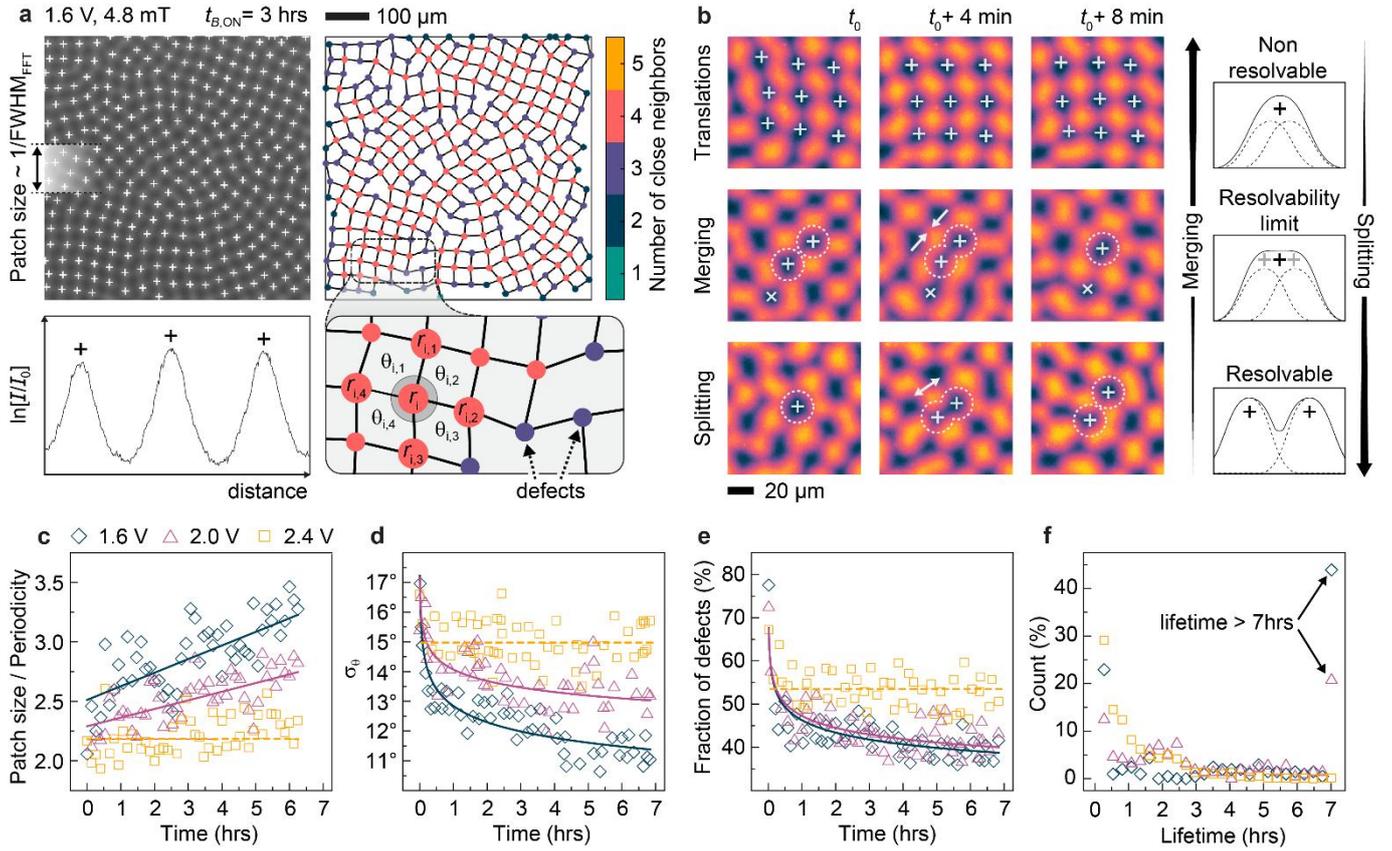

**Fig. 4 | Active dynamics. a.** Example of the relative neighborhood graph and details on the peak recognition and arrangement. **b.** Showcase of the phenomena associated with the activity in the pattern captured for the 2.4 V dataset. The color scale has been chosen to make the behavior more readable. **c.** Estimation of the ratio between the patch size as the inverse of the FWHM of the FFT peak and pattern periodicity. **d.** Standard deviation of the angle distribution in the pattern as a function of time. **e.** Fraction of defects (triangular coordination vertices) in the pattern as a function of time. **f.** Graph of the lifetime of the peaks which trajectory ended during the last 30 minutes of the measurement.

## Dynamics analysis and a minimal model capturing steady-state pattern dynamics

The observed overall complex dynamics of peak splitting, merging and collective translation that results in a continuous generation of defects in the pattern is quantified via differential dynamic microscopy (DDM) that allows both intuitive visualization and quantitative characterization of the pattern changes over time (Fig. 5a, b). We hypothesize that, in this regime, active behavior is linked to the generation of electrokinetic convective flows in the liquid, the strength of which depends on the applied voltage, as shown in our previous study[18]. Convective flow cells confined between plates, either generated by heat[20] or electric fields[21], are known to assemble into a hexagonal pattern. This implies that two competing pattern forming processes exist in our system, namely, the magnetic field induced instability (preferring square pattern) and electric field induced convection (preferring hexagonal pattern). As the competing pattern formation processes both involve instabilities, merely a superposition of the two driving forces cannot be expected to explain the resulting patterns. Indeed, the simple continuum theory presented alongside the particle-based simulations (Supplementary Notes 1 and 2) highlight the interplay of the electric and magnetic field in the pattern formation process (see Eq. (1.26) and Fig. S4, respectively).

In a rational attempt to understand the resulting pattern formation from this coupling, we formulate a Swift-Hohenberg (SH) model that captures the interaction of the two driving forces. The NP concentration (represented by an order parameter $\phi_1$) is modelled by an equation promoting a square lattice geometry, while the effects of the electroconvective flow via another order parameter $\phi_2$ promoting a hexagonal lattice geometry (Fig. 5c):

$$\partial_t \phi_1 = [\epsilon_1 - (q_1^2 + \Delta)^2]\phi_1 + \alpha_1 \phi_1^2 - \beta_1 \phi_1^3 - \gamma_1 \phi_1 \Delta^2 \phi_1^2 + (c_v + c_n)\phi_2 \tag{1}$$

$$\partial_t \phi_2 = [\epsilon_2 - (q_2^2 + \Delta)^2]\phi_2 + \alpha_2 \phi_2^2 - \beta_2 \phi_2^3 + (c_v - c_n)\phi_1, \tag{2}$$

where $\epsilon_i$ controls the distance from the onset of each instability, $(q_i^2 + \Delta)^2$ sets the pattern wavenumber, $\alpha_i, \beta_i, \gamma_i$ are the prefactors of the non-linear terms that set the pattern geometry, and the terms $(c_v \pm c_n)$ capture the interaction of the pattern geometries (see Supplementary note 3 for details). While this coupled continuum model with best achieved parametrization qualitatively reproduces the active dynamics of the experiment (Fig. 5c, Movie 7), the model also exhibits large scale collective translation of the peaks absent in the experiments. A more robust model can be obtained by coupling a random advection field $\boldsymbol{v}$ (with Langevin-like dynamics, driven by spatially correlated noise $\boldsymbol{\eta}$ and decay in time with a rate $\zeta$) to $\phi_1$ as

$$\partial_t \phi_1 = [\epsilon_1 - (q_1^2 + \Delta)^2]\phi_1 + \alpha_1 \phi_1^2 - \beta_1 \phi_1^3 - \gamma_1 \phi_1 \Delta^2 \phi_1^2 + \boldsymbol{v} \cdot \nabla \phi_1, \tag{3}$$

$$\partial_t \boldsymbol{v} = -\zeta \boldsymbol{v} + \boldsymbol{\eta}. \tag{4}$$

This random advection continuum model hence incorporates effect of $\phi_2$ on $\phi_1$ phenomenologically. For the correlation length of the noise, we used the pattern periodicity / $2\pi$, which is a natural choice to cause localized advection of peaks (Fig. S9).

Analysis of the experimentally observed pattern dynamics using DDM revealed that pattern characteristic persistence time ($\tau_q$) decreases linearly as a function of the voltage (Fig. 5e, f). Similar scaling was observed in simulated order parameter fields $\phi_1$ as a function of the negative decay rate -$\zeta$ (Fig. 5g, h). Hence, there is a linear relationship between the unitless simulation time and real time, and a linear relationship between $U$ and $\zeta$ as $\zeta \propto (U - U_s)$, where $U_s$ is a constant to account for the offset in voltage. This allows us to scale the dimensionless simulation data to match the time and length scales in the experiments (at the onset of activity – i.e., when activity becomes measurable). Using again vertex analysis as done earlier (Fig. 4), we define, for both the experimental data and the scaled simulation data, the activity rate $\rho = \frac{n_+ + n_-}{2n}$, where $n_+$ is the generation rate of new peaks, $n_-$ is the annihilation rate of existing peaks, and $n$ is the total number of peaks within the observation window. Good agreement is found between experiments and simulations (Fig. 5i). Similarly, agreement is found between the peak translation speed (Fig. 5j). These agreements indicate that the random advection continuum model captures the essential pattern formation and evolution features of the experimental system.

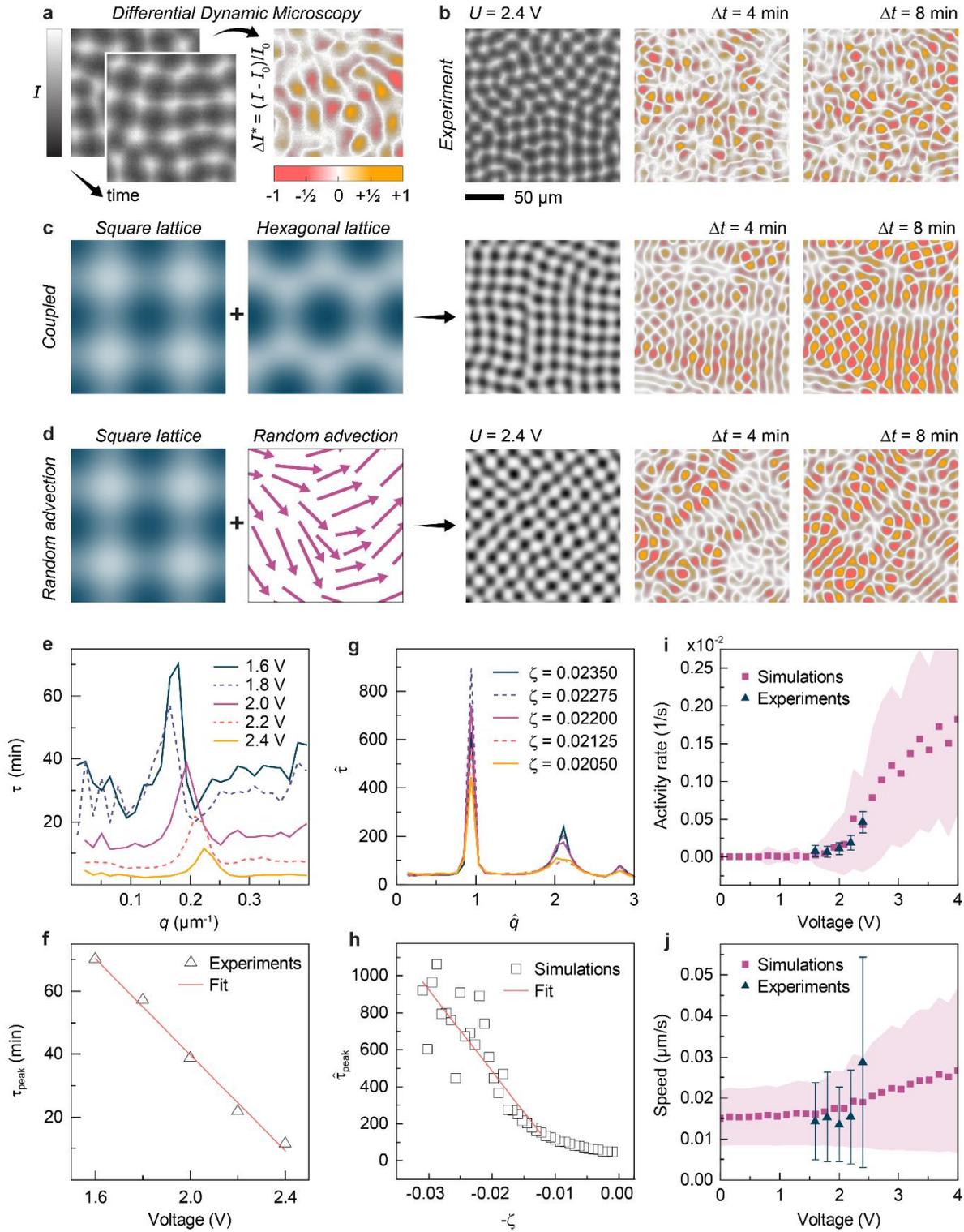

**Fig. 5 | Dynamic analysis and minimal continuum models. a.** Basic concept of the DDM analysis. The resulting intensity difference images $\Delta I$ have been normalized via $\Delta I^* = \Delta I / I_0$ to highlight the changes (graph color saturation at -1/2 and +1/2). **b.** Example of the DDM analysis applied to the experimental images of the 2.4V dataset with $t_0 \sim 3$ h. **c.** Schemes of the expected $\phi$-values from the coupled continuum model with a square and hexagonal lattice geometry (left) and an example of the DDM analysis for the coupled continuum model with rescaled time (right). **d.** Schemes of the expected $\phi_1$-values from the random advection continuum model resulting in a square lattice and the direction of the advection in space (left) and an example of the DDM analysis for the random advection continuum model with rescaled time (right). **e.** DDM analysis of the experimental data in the annealing timeframe. **f.** Plot of the $\tau$ maxima values of the peak as a function of the voltage. **g.** DDM analysis of selected simulations data chosen as their behavior was the most similar looking to the experiments. **h.** Plot of the $\hat{\tau}$ maxima values of the peak as function of the parameter $\zeta$. **i.** Steady-state activity rate and **j.** Steady-state peak speed of the vertices as a function of the voltage applied. The presented values are average and standard deviation, for experiments between hours six and seven from the onset of the pattern, and for simulations, for the last 500 frames, corresponding to ~4.26 h.

## Conclusions and outlook

In summary, we have presented an experimental demonstration of an active version of the classic Rosensweig pattern, where the pattern does not reach a stationary minimum-energy structure but rather continuously fluctuates. We have interpreted the experimental results within a scaling theory approach, as well as in terms of discrete active particles, and supported the findings by simulations of minimal non-equilibrium continuum models, as well as particle-based simulations. Our work suggests that it may be possible to "activate" magnetically and otherwise driven equilibrium pattern forming systems to yield minimalistic non-equilibrium systems that mimic the features of far more complicated systems, including living ones.

## Methods

***Materials:*** Iron (III) chloride hexahydrate (FeCl$_3$ · 6H$_2$O, ≥ 99%, Sigma-Aldrich), iron (II) sulfate heptahydrate (FeSO$_4$ · 7H$_2$O, ≥ 99%, ACS reagent, Sigma-Aldrich), ammonium hydroxide (NH$_4$OH, 28-30%, ACS reagent, Sigma-Aldrich), oleic acid (C$_{18}$H$_{34}$O$_2$, 90% technical grade, Sigma-Aldrich), acetone (C$_3$H$_6$O, ≥ 99.8%, Fisher Scientific), toluene (C$_6$H$_5$CH$_3$, ≥ 99.7%, ACS reagent, Sigma-Aldrich), *n*-dodecane (C$_{12}$H$_{26}$, 99%, anhydrous, Acros Organics) and docusate sodium salt (C$_{20}$H$_{37}$NaO$_7$S, AOT, ≥99%, anhydrous, Sigma-Aldrich) were used as received.

***Synthesis and characterization of the stock dispersion of iron oxide nanoparticles in toluene:*** Synthesis and characterization of the iron oxide nanoparticles was done as previously[17] with minor modifications. In brief, the NPs were synthesized using the Massart coprecipitation method[22] with oleic acid as stabilizing surfactant and toluene as carrier fluid. The mass fraction of the nanoparticles in the stock dispersion was determined by measuring the weight loss during evaporation of the carrier solvent using an analytical balance (Ohaus Pioneer PX224). The corresponding volume fraction was calculated by using the density of the ferrofluid. The density of the toluene based ferrofluid was determined as follows, following the procedure validated before[6]: 100 μl of ferrofluid were pipetted (Eppendorf Multipipette E3x equipped with Combitips advanced 0.1 ml) and weighed with an analytical balance. Each measurement was performed at room temperature (22 ± 1 °C) in quintuplets. The density of deionized (DI) water was measured in parallel to validate the obtained values ($\rho_{\text{water}}$: 0.99 ± 0.01 g ml$^{-1}$). The final averaged value of the density of the toluene based ferrofluid was calculated to be: 0.910 ± 0.009 g ml$^{-1}$. The results of the characterization are summarized in **Table 1**.

**Table 1**. Physical properties of the stock dispersion of iron oxide nanoparticles in toluene.

| **Measured property** | **Measured value** |
|---|---|
| Density of the stock dispersion | 0.910 ± 0.009 g ml$^{-1}$ |
| Mass fraction of nanoparticles in the stock dispersion | 8.56 ± 0.04 % |
| Volume fraction of nanoparticles in the stock dispersion | 2.58 ± 0.04 % |

***Preparation and general characterization of the electroferrofluid:*** Preparation and characterization of the electroferrofluid was done as previously[17] with small modifications. Briefly, five 2 ml vials were filled with 200 μl of the stock dispersion of iron oxide nanoparticles in toluene and with 200 μl of a 300 mM solution of AOT in dodecane. The vials were left open in a fume hood, under ambient temperature and humidity, for ca. 80 hours to let the toluene evaporate. At the end of the process, the content of the five vials was mixed in a single vial. The density of the electroferrofluid was determined as for the stock ferrofluid. The results of the characterization of the electroferrofluid are summarized in **Table 2**.

*Magnetic characterization of the electroferrofluid:* Magnetic properties of the stock electroferrofluid were measured using a vibrating sample magnetometer (QuantumDesign PPMS VSM DynaCool). First, the magnetic field applied using the VSM was calibrated by using a palladium reference sample (Quantum Design QDS-C210A) with a mass of 262 mg and magnetic responsivity of $1.3755 \cdot 10^{-2}$ emu/T (Fig. S1a). After that, a VSM powder sample holder (P125E) was filled in the bottom with a mass of 7.32 mg of electroferrofluid measured with analytical balance. The capsule was then sealed by thermally bonding it with another sample holder to avoid evaporation of the electroferrofluid during the measurement in vacuum inside the VSM. The VSM measurement of the filled capsule (Fig. S1b) was performed using the same protocol used for the Palladium calibration measurement. Finally, a glass capillary of 0.22 mm of diameter (HIRSCHMANN ringcaps® 10 μl) cut at 4 mm length was filled with a small amount of electroferrofluid (< 1 μl). The ends of the capillary were sealed with molten wax (PELCO® Quickstick 135) to avoid evaporation. The high aspect ratio of the sample guaranteed a negligible demagnetizing effect in the whole range of the measurement. The final magnetization curve was obtained by repeating the protocol used for the first two preliminary measurements, calibrating the applied field with the result from the palladium sample measurement and the saturation magnetization with the result of the capsule measurement (Fig. S1c). The datapoints were interpolated with a grade 1 polynomial close to the origin to measure the magnetic susceptibility. The final values for the magnetic properties of the material are summarized in **Table 2**.

**Table 2.** Physical properties of the stock electroferrofluid.

| **Measured property** | **Measured value** |
| --- | --- |
| Density | $0.844 \pm 0.008$ g ml$^{-1}$ |
| Volume fraction of AOT | $11.81 \pm 0.04$ % |
| Volume fraction of nanoparticles | $2.51 \pm 0.04$ % |
| Saturation magnetization | $4.078 \pm 0.005$ kA m$^{-1}$ |
| Magnetic susceptibility | $0.095 \pm 0.002$ |

*Microscopy:* Observation of the sample was performed using a custom build microscope (Fig. S2c) with main components from the Thorlabs Cerna® series (Thorlabs CSE2100, Thorlabs CEA1400, Thorlabs WFA4102, Thorlabs CSA1001). The microelectrode cell was illuminated in transmitting light using a white LED light source (Thorlabs MNWHL4). The objective used for the imaging is a 20x infinite conjugated microscope from Nikon (Nikon 20x objective TU Plan Fluor). The images were collected using a 5 MP grayscale digital camera (Ximea MC050MG-SY-UB) equipped with a heat sink (MECH-25MMHEATSINK-KIT). The camera control was performed using a software provided by the camera manufacturer (Ximea xiCamTool 4.28). Image length scale was calibrated using a calibration target (Thorlabs R1L3S2P). Due to the long-time length of the experiments performed, the light source activation was synchronized with image collection using the in-built sync signal of the digital camera and an 8-pin cable (CBL-702-8P-SYNC-5M0) connected to the LED power supply (Thorlabs LEDD1B). Each frame saved was the result of the average over a burst of 10 frames collected in a time window of 50 ms. Analysis and elaboration of the experimental images has been performed with MATLAB[23] or ImageJ/Fiji[24,25]. Grayscale values in the images shown in Fig. 1, 4, and 5 have been adjusted to help visualization.

***Cell preparation and application of external voltages and B-fields:*** A specific alloy of solder (Cerasolzer GS182) was bonded to the ITO coating of the microelectrode cell (Instec D400A200uNOPI) using an ultrasonic soldering iron (MBR electronics USS-9210 Ultrasonic Soldering System). Metal wires were then connected to the solder bonded with ITO using a standard soldering iron (Weller T0053298699) and solder (Multicomp 509-0600). The cell was filled using a micropipette (Eppendorf Research Plus) and sealed with molten wax (PELCO® Quickstick 135) to avoid evaporation. The voltage was applied on the microelectrode cell using a high resistance meter (Keysight B2987A) or, alternatively, for the longer sets of measurements, a standard DC bench power supply (Multicomp Pro MP710087). The magnetic field was applied as before[17] using a pair of electromagnet coils (GMW 11801523 and 11801524) connected to DC power supply (BK Precision 9205). Both power supplies used for the application of voltages and for the powering of the coils were controlled using a software developed with LabView that allows synchronization with the image collection.

***Periodicity analysis and pattern structure:*** The pattern periodicity was analyzed using a combination of ImageJ/Fiji, plugins, and macros and scripts developed in MATLAB (Fig. S3). First, the positions of the maxima are located using the FindRegionalMax command after which the Delauney triangulation combined with a Relative Neighborhood Graph[26] were used to determine the distances between the neighboring maxima. The average distance between the peaks and its standard deviation were calculated from the resulting peak-to-peak distance distribution. The analysis outcome was confirmed by comparing the outcome against the Fast Fourier Transform (FFT) of each microscopy frame in ImageJ/Fiji software. The polar-angle average profile of each FFT image was obtained using the ImageJ/Fiji Radial Profile Plot plugin. Each obtained FFT radial spectrum was analyzed by a MATLAB script to locate the position and HWHM of the most frequent length scale in the pattern. The Relative Neighborhood Graph and the FFT approach led to periodicity results that consistently fall into each other error bars. OriginPro[27] was used for data plotting, and also for the fitting of the trend models to the datasets.

The optimization problem to collapse the data to obtain the plot in the inset of Fig. 3a was implemented with a MATLAB script (Fig. S5).

The pattern geometry was analyzed for each frame in the time series datasets using a MATLAB script. The pattern vertex angles and the number of defects in the pattern were calculated using the Relative Neighborhood Graph[26]. Pattern dynamics were quantified by a MATLAB script that tracked the vertices of the pattern in time frame by frame, keeping count of the peak annihilation and peak generation events. This allows for the measurement of the peak speed (Fig. 5 and Fig. S8). The annihilation and generation rates were calculated by dividing for each frame the number of annihilation and generation events by the total number of vertices. Sample drifting was accounted for by subtracting the average movement of all vertices in each analysis frame from the individual peak velocities.

*IV-curve measurement:* the current-voltage response (IV-curve) of the electroferrofluid in the cell was measured by the current flowing in the cell with a high resistance meter (Keysight B2987A). In brief, the measurement was by increasing step-wise the voltage applied every 30 s and measuring the average value of the current in the last 2 s of the step (Fig. S5).

*Amplitude analysis:* The "amplitude" of the pattern can be estimated by measuring the light intensity in the location of maximum brightness in the pattern (Fig. S6) and taking advantage of the Beer Lambert law for light absorption as $\sim \ln[I(x_{\max}, y_{\max})/I_0]$ where $I(x_{\max}, y_{\max})$ is the light intensity of the maximum values in the images corresponding to the areas with less NPs and $I_0$ is the light intensity of the uniform concentration. The analysis of the NPs displacement in the cell was performed using a combination of ImageJ/Fiji commands, plugins and scripts developed in MATLAB (Fig. S6). First, the microscopy images were smoothed to reduce the single pixel noise by replacing each pixel with the average of its $3 \times 3$ neighborhood (Smooth command in ImageJ/Fiji). Next, the maxima of the image were located framewise with the Find Maxima Command in ImageJ/Fiji. The average value and the standard deviation for all maxima were extracted from the obtained data using a MATLAB script. OriginPro was used for plotting and for fitting of the trend models for the datasets. A brief note regarding this analysis: taking advantage of the Beer-Lambert law in this way implies that we are integrating the absorption of light over the cell thickness effectively averaging the effect of the local concentration of NPs $c(x, y, z)$ in the $z$ direction. A complete reconstruction of the 3D profile of the NP concentration, and consequently a better definition of the amplitude of the pattern, would require fully fluorescent magnetic NPs and confocal microscopy. Unfortunately, such NPs are still eluding the research community.

*DDM:* The data of both the microscopy images and time series obtained from simulations were analyzed using a differential dynamic algorithm[28,29]. This calculates characteristic times $\tau_q$ at different wavenumbers $q$. The characteristic times $\tau_q$ are obtained from analyzing the difference signal of the pixel values $I$, $\Delta I(x, y; \Delta t) = I(x, y; t = \Delta t) - I(x, y; t = 0)$ as a function of the time delay $\Delta t$. The approach captures the time and length scales at which the observed pattern evolves. The length scale of the pattern is easily obtained from analyzing the Fourier transform $F_{\Delta I}(q_x, q_y; \Delta t)$ of the difference signal $\Delta I(x, y; \Delta t)$. An azimuthal average gives the one-dimensional power spectrum $|F_{\Delta I}(q; \Delta t)|^2$, where $q = \sqrt{q_x^2 + q_y^2}$. For each wave number, $|F_{\Delta I}(q; \Delta t)|^2$ first increases with $\Delta t$ before saturating. The characteristic time $\tau_q$ of each wavenumber mode is obtained by fitting

$$|F_{\Delta I}(q; \Delta t)|^2 = A(q)\left[1 - \exp\left(-\frac{\Delta t}{\tau(q)}\right)\right] + C(q), \tag{5}$$

for each wavenumber. $A(q)$ and $C(q)$ encode the wavelength dependent relationship between intensity and concentration fluctuations, and camera noise, respectively. However, for fixed $q$ they can be treated as adjustment parameters. The thus obtained characteristic times $\tau_q$ show a clear peak at the observed pattern's wavenumber. OriginPro was used for plotting the results of the geometry and dynamic analysis.

*Computational models and theory:* A continuum model for nanoparticle distributions and scaling analysis of pattern onset was formulated (Supplementary Note 1). Standard Brownian dynamics[30] particle-based simulations of a system including only neutral nanoparticles with permanent magnetic dipole moments were performed to examine NP distribution and external magnetic field competition (Supplementary Note 2). Active continuum model simulations were performed to capture features in the active behavior arising from the nonlinear interaction of the electric and the magnetic field effects (Supplementary Note 3).

## Supplementary Materials

Notes 1-3

Figures S1-S9

Movies 1-9

## Data and materials availability

Raw image and movie data files and processing codes are available in a dataset published in Zenodo: 10.5281/zenodo.12607604 (data will be made public after acceptance of the manuscript).

## Code availability

The code used for the simulations performed is available in a dataset published in Zenodo: 10.5281/zenodo.12607604 (data will be made public after acceptance of the manuscript).

## Acknowledgements

This work was supported by Academy of Finland through its Centres of Excellence Programs (2022-2029, LIBER) under projects no. 346111 (M.S.), 346112 (J.V.I.T.), 364205 (M.S.), 364206 (J.V.I.T), Academy of Finland through projects no. 342038 (C.R.), 359180 (M.S.), 342116 (J.V.I.T.), 316219 (J.V.I.T.), and European Research Council (ERC) under the European Union's Horizon 2020 research and innovation programme (grant agreement no. 803 937) (J.V.I.T.). M.P. Haataja was supported by the National Science Foundation through the Princeton University (PCCM) Materials Research Science and Engineering Center DMR-2011750 and A.S. by the Swiss National Science Foundation under the project no. P500PT_206916. Computational resources by CSC IT Centre for Finland, the Aalto Science-IT project, and RAMI – RawMatters Finland Infrastructure are also gratefully acknowledged. We acknowledge CSC for awarding this project access to the LUMI supercomputer, owned by the EuroHPC Joint Undertaking, hosted by CSC (Finland) and the LUMI consortium through a CSC project call.

## Author Contributions

C.R. synthesized the nanoparticles, prepared the electroferrofluids, designed and constructed the experimental setups, characterized the samples, acquired the final microscopy data, analyzed the experimental data, wrote the first draft of the manuscript including the figures and movies. M.P. Haataja and E.S. developed the theoretical models. M.P. Holl designed and performed the active continuum model simulations and the DDM analysis and cured the dynamics data match with the experimental data. A.S. designed and performed the particle-based simulations and matched the findings with the continuum models and experimental data. F.S. performed preliminary experiments and helped with ferrofluid synthesis. M.S. and M.P. Haataja supervised the work of M.P. Holl and E.S. J.V.I.T. led the project and supervised experiments and data analysis. All authors participated in the editing of the final manuscript.